\documentclass[onecolumn,10pt]{IEEEtran}
\usepackage{cite,amsmath,amssymb,amsfonts,color}
\usepackage[dvips]{graphicx}
\usepackage{setspace}
\usepackage{amsmath,subfigure,amssymb,amsfonts,color,epsfig}
\usepackage{mathrsfs}
\usepackage{xcolor}
\usepackage{lipsum}
\usepackage{caption}
\usepackage{epstopdf}
\usepackage{mathrsfs}
\usepackage{diagbox}

\usepackage{multirow}
\usepackage{multicol}
\DeclareMathOperator*{\argminA}{arg\,max} % Jan Hlavacek
\usepackage{algorithmic}
\usepackage{algorithm}
\usepackage{amsthm}

\usepackage{xcolor}

\begin{document}
\title{Detecting Changes in Time Series Data using Volatility Filters}
%%%%%%%%%%%%%%%%%%%%%%%%%%%%%%%%%%%%%%%%
\author{Alireza Ahrabian, Nazli Farajidavar, Clive Cheong-Took and Payam Barnaghi\\%,~\IEEEmembership{Fellow,~IEEE}\\
\thanks{The authors Alireza Ahrabian and Payam Barnaghi are with the Department of Electrical and Electronic
Engineering, Institute for Communication Systems at the University of Surrey (email:
\{a.ahrabian, p.barnaghi\}@surrey.ac.uk). \newline
\indent Clive Cheong-Took is with the Department of Computer Science at the University of Surrey  (email:c.cheongtook@surrey.ac.uk).
\newline 
\indent Nazli Farajidavar is with the Computational Health Informatics laboratory at the University of Oxford (email:nazli.farajidavar@eng.ox.ac.uk).}} \vspace{-6cm}
\maketitle

\begin{abstract}
\indent This work develops techniques for the  sequential detection and location estimation of transient changes in the volatility (standard deviation) of time series data. In particular, we introduce a class of change detection algorithms based on the windowed volatility filter.  The first method detects  changes by employing a convex combination of  two such filters with differing window sizes, such that the adaptively updated convex weight parameter is then used as an indicator for the detection of instantaneous power changes. Moreover, the proposed adaptive filtering based method is readily extended to the multivariate case by using recent advances in distributed adaptive filters, thereby using cooperation between the data channels for more effective detection of change points.  Furthermore, this work also develops a novel change point location estimator  based on the differenced output of
the volatility filter. Finally, the performance of the proposed
methods were evaluated on both synthetic and real world data.
\end{abstract}
\vspace{-2mm}
\section{Introduction}
The piecewise segmentation of time series data into intervals with similar
stochastic characteristics is often important in the exploratory analysis of data and is generally referred to as change detection. With rapid growth in sensing technologies, there is a need for effective and adaptive methods to identify and analyse activities in  time series data \cite{Benveniste83}. In particular the partitioning  of data into segments of  constant volatility (variance/standard deviation)  has  interesting applications for fields ranging from the segmentation of biomedical data \cite{Sejdic09} to audio segmentation \cite{Desobry05}\cite{Obrecht88}. 
\\
\indent Algorithms proposed for estimating the transition times for data points with transient changes in the variance are generally divided into two categories: i) Offline algorithms that use the entire data set so as to estimate the location of the change points and ii) online (sequential) methods, where data is only available up to the current time sample being analysed. Early research primarily focused on developing statistical significance based change detection algorithms, where in particular the  work in \cite{Inclan94} proposed a method based on the cumulative sum of squares in combination with a statistical significance test  to provide an estimate of  change points. More recently, algorithms based on Markov Chain Monte Carlo (MCMC) techniques for obtaining estimates of the change point locations have been developed \cite{Punskaya02}\cite{Wang03}\cite{Lavielle98}. However, such methods are computationally expensive and process the data offline. More computationally efficient block based dynamic programming \cite{Killick12} and Bayesian methods  \cite{Lavielle05} have been proposed, however, they have limited use for online detection of change points. It should be noted, that block based multivariate change point detection algorithms have also been developed. In particular the work in \cite{Xuan07} developed an efficient Bayesian approach, while the work in \cite{Bleakley11} used the group fused LASSO. 
\\
\indent Many real world practical change detection problems require sequential algorithms that are not only capable of detecting with the smallest possible delay the change  volatility, but also  providing a good estimate of the change point location. The work in \cite{Brandt83} introduced a method based on the likelihood ratio test that includes a simple threshold and test statistic for both  detecting and estimating the location of the change in the volatility  of the time series data. Furthermore, the work in \cite{Wang03} proposed a threshold independent sequential change detection method using the minimum description length (MDL), while an  accurate online kernel-Support Vector Machine (k-SVM) based method  was proposed in \cite{Desobry05}. However, such methods often have drawbacks, including: high computational complexity,  low accuracy of change point location estimate and high detection latency. 
\\
\indent We introduce a class of change detection and location estimation algorithms based on the windowed volatility filter \cite{Alexander08}.  The first proposed algorithm seeks to detect such transient power/volatility variations as rapidly as possible. This is achieved by employing a convex combination of two such filters, where a state change is detected by selecting the appropriate filter  given the statistics of the data points. The proposed method is designed based on  the trade off that exists between the  steady state accuracy and convergence speed of volatility filters. By using an adaptive filter  \cite{Zhang06}\cite{Garcia106}, the proposed method seeks to overcome this trade off by updating the convex weight parameter that combines the respective windowed volatility filters, thereby providing an indicator (in the form of the convex weight parameter) for detecting changes in the volatility of the  data. An advantage of the use of adaptive filters for the change point detection problem is a  multivariate extension of the proposed algorithm that is readily developed by using recent advances in distributed signal processing \cite{Sayed08}. Furthermore, this work analyses the behaviour of the volatility filter weights on the performance of the change detection algorithm, thereby providing a means of selecting such weights for improved performance.  The second proposed method provides an estimate  for the change point location. This is achieved by proposing a new estimator based on the differenced output of the volatility filter. The performance of both the proposed change detection and location estimation techniques  are verified on both synthetic and real world data sets. 
 
\begin{figure}[t!]
\begin{center}
\includegraphics[width=0.5\columnwidth]{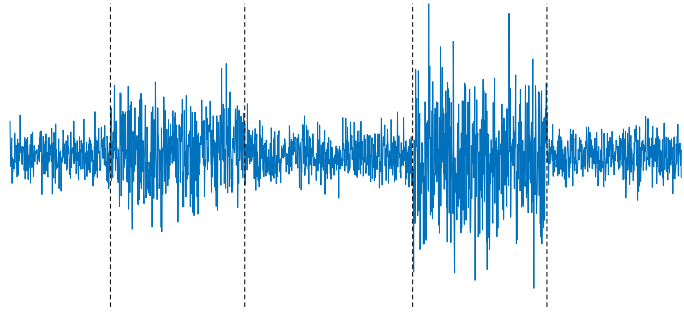}
\caption{Time series data with instantaneous power changes, where the vertical dashed lines show the location of the transition points.}\label{fig:ChangeExample}
\end{center}
\end{figure}
\vspace{-3mm}
\section{Problem Outline}
Given a sequence of data points $\{x_{1},x_{2},x_{3},\dots,x_{t}\}$ drawn from a zero mean Gaussian distribution with  piecewise constant variance;  there exists an unknown sequence of transition time instances $\{\tau_{1},\tau_{2},\dots,\tau_{t}\}$ at which a change  in the variance occurs (an example is shown in Fig.~\ref{fig:ChangeExample}). Accordingly, this work seeks to address the following problems: 1)  detecting a change in the volatility of time series data with the shortest possible latency; 2)  providing an accurate online estimate of the location of the transition times $\{\hat{\tau}_{1},\hat{\tau}_{2},\dots\,\hat{\tau}_{t}\}$. 
\vspace{-1mm}
\section{Proposed Algorithms}
This work proposes   volatility filtering based change detection and location estimation algorithms given a zero mean set of data points, $\boldsymbol{\text{x}}=[x_{1},x_{2},x_{3},\dots,x_{N}]$; where an estimate of  volatility   is given as the output of the volatility filter, $\hat{\sigma}_{p}(t)$. In particular the volatility filter is defined as the finite impulse response (FIR) filtered estimate  of the past $T_{p}$  squared data points, that is
\begin{equation}
\begin{aligned}
\hat{\sigma}^{2}_{p}(t) &= w_{p,1} x^{2}_{t}+...+w_{p,T_{p}}x^{2}_{t-T_{p}+1}\\
&=\boldsymbol{\text{w}}_{p}^{T}\boldsymbol{\text{x}}_{p}
\end{aligned}
\label{varFiltP}
\end{equation}
\noindent where $\boldsymbol{\text{w}}_{p}=[w_{p,1},...w_{p,T_{p}}]$ correspond to the set of volatility filter weight coefficients with $\sum_{i=1}^{T_{p}}w_{p,i}=1$ (for the unbiased estimate of variance this equality is approximately satisfied for large samples) and $\boldsymbol{\text{x}}_{p}=[x^{2}_{t},...,x^{2}_{t-T_{p}+1}]$. By using the volatility filter as defined in \eqref{varFiltP}, we first develop an adaptive filtering based change detection algorithm that utilises a convex combination of two such distinct volatility filters. We also provide analysis for selecting appropriate  filter weights for the respective volatility filters. The second proposed method is a change point location estimator that is based on the output of the differenced volatility filter. 
\subsection{Adaptive Filtering Based Change Detection Method}
\begin{figure}[t!]
\begin{center}
\includegraphics[width=0.5\columnwidth]{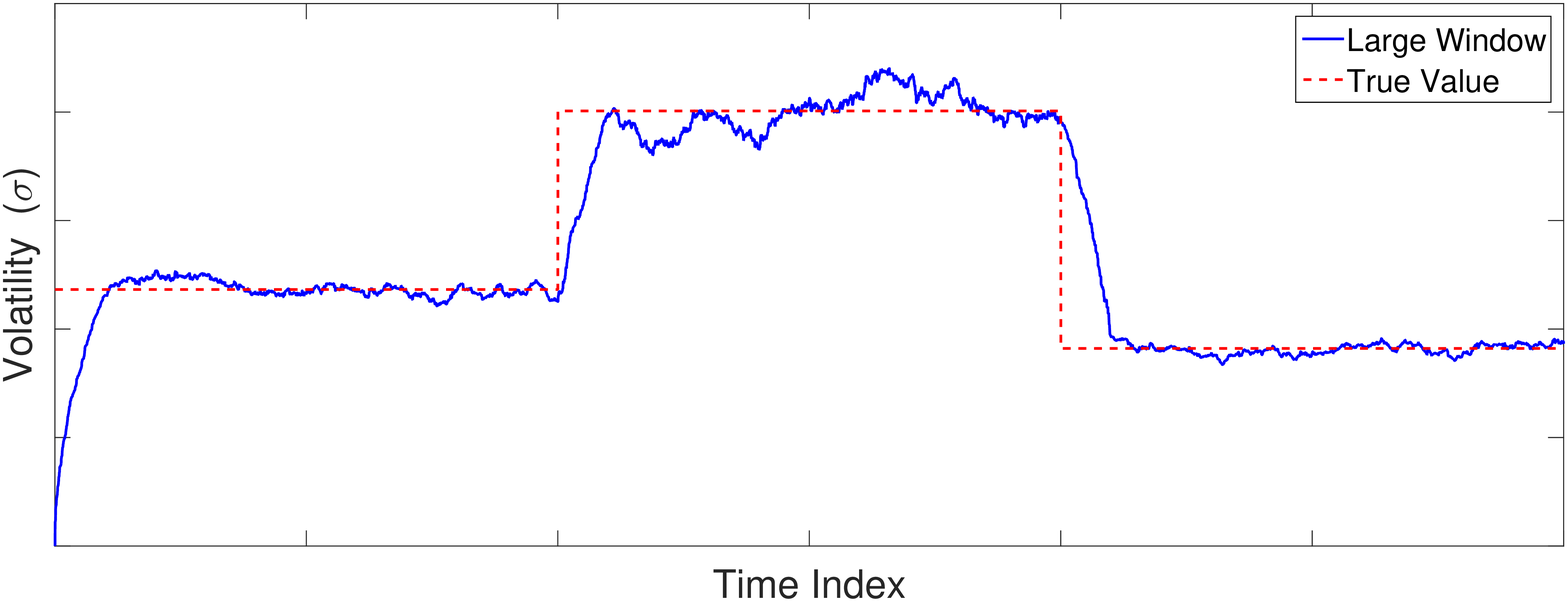}
\includegraphics[width=0.5\columnwidth]{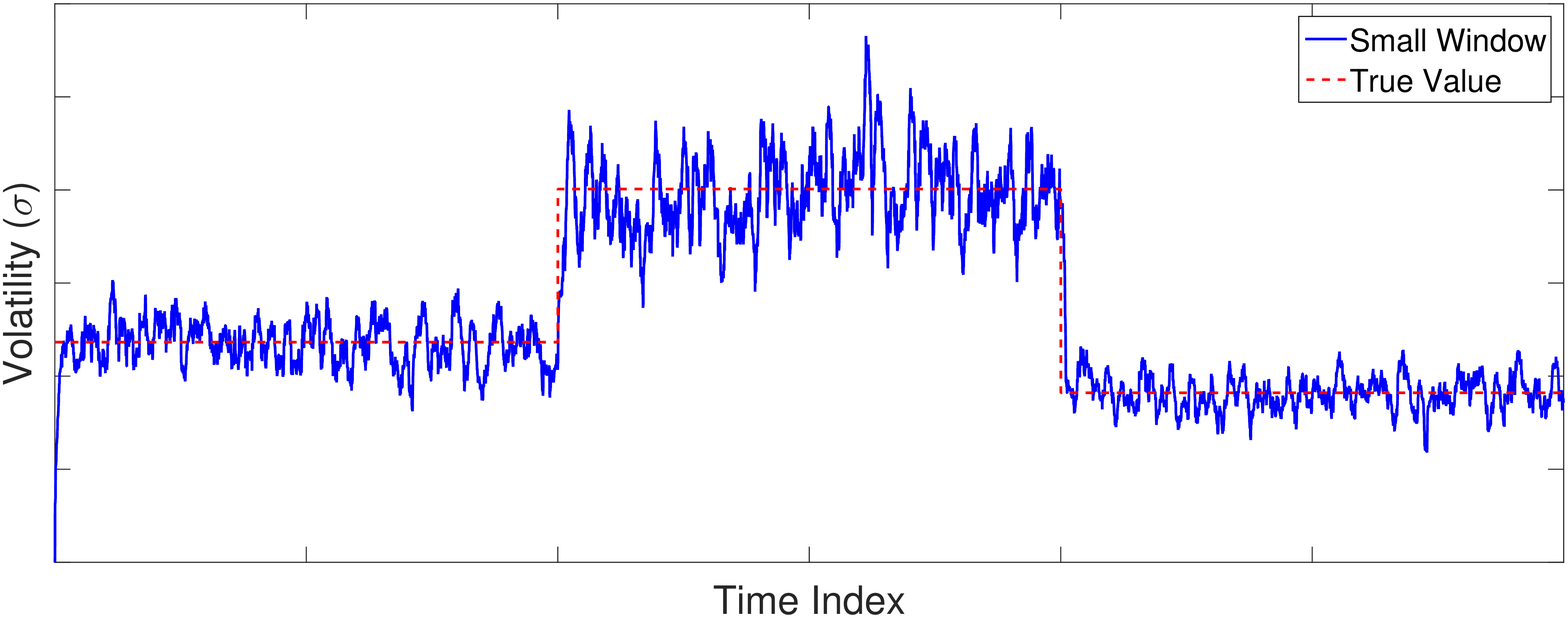}
\caption{Figure showing the estimated volatility for time series with  instantaneous power changes.  (Upper panel) Shows the output
of the `slow' and (lower panel) the output of the `fast' volatility filter.
 }\label{fig:SlowExample}
\end{center}
\end{figure}

This section presents an adaptive filtering and volatility filtering based change detection algorithm. This is achieved by using an adaptive filter \cite{Jelfs08}\cite{Zhang06}\cite{Garcia106} to track changes in the instantaneous standard deviation using a  convex combination  of two volatility filters with different window sizes. Where a volatility filter with a small window size is able to effectively track transient changes in the time series volatility; however, in the steady state the estimate  in the volatility of the signal is not accurate (please see Fig.~\ref{fig:SlowExample}). While a volatility filter with a large window size has stable steady state estimates of volatility, however, it is not able to track changes effectively (please see Fig.~\ref{fig:SlowExample}). By adaptively estimating the convex weight parameter, such that if a significant transient change has occurred then volatility filter with a small window size provides a more optimal (least squared sense) solution than the volatility filter with the large window size. Implying that by tracking the estimate of the convex weight parameter, it can be determined whether or not a change has occurred. Furthermore, the proposed change detection method can  be extended to  data sets drawn from multiple sensors, by using an adaptive filtering architecture that utilises  cooperative strategies \cite{Sayed08} to enhance both convergence and steady state accuracy of the convex weight parameter.  
\\
\indent Given a set of data points $\boldsymbol{\text{x}}$, consider the following volatility filters
\begin{equation}
\begin{aligned}
\hat{\sigma}^{2}_{f}(t) &= w_{f,1}x^{2}_{t}+...+w_{f,T_{f}}x^{2}_{t-T_{f}+1}\\
&=\boldsymbol{\text{w}}_{f}^{T}\boldsymbol{\text{x}}_{f}\\\\
\hat{\sigma}^{2}_{s}(t) &=w_{s,1} x^{2}_{t}+...+w_{s,T_{s}}x^{2}_{t-T_{s}+1}\\
&=\boldsymbol{\text{w}}_{s}^{T}\boldsymbol{\text{x}}_{s}
\end{aligned}
 \label{voltastslow}
\end{equation}
where $\hat{\sigma}_{f}(t)$ and $\hat{\sigma}_{s}(t)$ correspond\footnote{We now refer to the volatility filters with relatively small and large window sizes as `fast' and `slow' filters respectively, this is to reflect the convergence speed of the filters when transient changes occur in the signal volatility.} to the respective outputs of the `fast' and `slow' volatility filters, such that, $T_{f}<T_{s}$. The proposed estimate of the instantaneous volatility $\hat{\sigma}_{o}(t)$ is given by a convex combination of the `fast' and `slow' filters, that is
\begin{equation}
\hat{\sigma}_{o}(t) = \lambda(t)\hat{\sigma}_{f}(t) + (1-\lambda(t))\hat{\sigma}_{s}(t)
\end{equation}
where $\lambda(t)$ is updated adaptively using the  adaptive filter architecture outlined in \cite{Zhang06}\cite{Garcia06}. 
The squared error, $\text{E}\{e^{2}(t)\}$, required to be minimised is generally  given by the following 
\begin{equation}
\text{E}\{e^{2}(t)\} = \text{E}\{\big(\hat{\sigma}_{d}(t)-\hat{\sigma}_{o}(t)\big)^{2}\},
\label{CF}
\end{equation}
where $\hat{\sigma}_{d}(t)$ is defined as
\begin{equation}
\begin{aligned}
\hat{\sigma}^{2}_{d}(t) &= w_{1}x^{2}_{t+1}+...+w_{T_{d}}x^{2}_{t-T_{d}}\\
&=\boldsymbol{\text{w}}_{d}^{T}\boldsymbol{\text{x}}_{d}
\end{aligned}
 \label{volatilityDesired}
\end{equation}

\begin{algorithm}[t!]
%\textbf{Initialise:}
 %  \indent Set $s(1)=1$, significance level $k$ and volatility filter window size $T$.
\begin{algorithmic} 
\REQUIRE 
\STATE - Input signal $x_{t}$  
\STATE - Select: $\mu, T_{s}, T_{f}, T_{d}, \gamma$.
\STATE - Set: $\lambda(1)\leftarrow 1$, $t\leftarrow1$, $\rho\leftarrow0.001$, $T_{r}\leftarrow1.2T_{s}$.
\vspace{1mm}
\WHILE{$t< N$}
\STATE -  $\sigma_{o}(t) \leftarrow \lambda(t)\hat{\sigma}_{f}(t) + (1-\lambda(t))\hat{\sigma}_{s}(t)$.
\STATE -  $e(t)\leftarrow \hat{\sigma}_{d}(t)-\hat{\sigma}_{o}(t)$.
\STATE  - $\lambda(t+1) \leftarrow \lambda(t) + \mu(|\lambda(t)|+\rho u)e(t)(\hat{\sigma}_{f}(t) - \hat{\sigma}_{s}(t))$
\IF{$\lambda(t+1)\geq\gamma$ and $\text{Flag}(t)=0$}
\STATE - $s(t)=1$ in the interval $[t,t+T_{r}]$.
\STATE - $\mu\leftarrow \frac{\mu}{\text{Var}(x_{t})}$ in the interval $[t,t+T_{r}]$.
\STATE - $ \text{Flag}(t)=1$ in the interval $[t,t+T_{r}]$.
\ENDIF
\STATE - $t \leftarrow t+1$
\ENDWHILE
\end{algorithmic}
%\textbf{Output} 
\caption{\!\!: Adaptive Filtering Based Change Detection}\label{alg:AdaptiveFilteringBasedChange}
\end{algorithm}
\noindent where $T_{d}$ is the window length of the desired volatility filter.  The desired signal $\hat{\sigma}_{d}(t)$ provides a coarse estimate of the instantaneous volatility so as to enable  the   adaptive filter to  track changes in volatility; to this end,  a typical window length of   $T_{d}\approx 10$ is recommended.
It should be noted that $\lambda(t)\rightarrow1$ during a change in the volatility, that is, a higher weight is placed on the 'fast' filter. Moreover,  in the steady state (variance does not change) a higher weight will be assigned to on the `slow' volatility filter; however, during simulations we found large variations in the weight estimate $\lambda(t)$ when using the conventional least mean square (LMS) update. To this end, we propose to use  signed sparse-least mean squares (SS-LMS) algorithm \cite{Martin02}, that is the minimised cost function \eqref{CF} along with a prior distribution on the values taken by $\lambda(t)$. By using a variation of the SS-LMS algorithm we ensure that the estimate of $\lambda(t)$ has lower variability as it  converges to  the `slow' volatility filter. Accordingly,  the  update of, $\lambda(t)$, is given by\footnote{A hard threshold is applied so that $0\leq\lambda(t)\leq1$, that is,
\begin{equation*}
\lambda(t)     = \left\{ \begin{array}{ll}
1, & \lambda(t)\geq 1\\
0, & \lambda(t)\leq 0.
\end{array} \right. \label{sthreshold}
\end{equation*} }
\begin{equation}
\lambda(t+1) = \lambda(t) + \mu(|\lambda(t)|+\rho u)e(t)(\hat{\sigma}_{f}(t) - \hat{\sigma}_{s}(t))
\label{SS-LMS}
\end{equation}
where $\mu$ is the learning rate of the adaptive filter, $\epsilon$ is a positive constant for regularization, $\rho$ is a  parameter selected so as to reduce the variability of $\lambda(t)$ and $u$ corresponds to a Gaussian distributed random variable. The updated convex weight parameter is then mapped to the output of the binary state change sequence $s(t)$, that is,
$s(t) = \left\{ \begin{array}{ll}
1, & \lambda(t)\geq \gamma\\
0, & \lambda(t)< \gamma
\end{array} \right.$,
where $0\leq\gamma\leq1$, is a threshold and $s(t)=1$ corresponds to a change in the volatility of the time series, while $s(t)=0$ corresponds to no change.  Algorithm 1 summarizes the proposed method referred to as the adaptive filtering based change detection (AFCD). It should be noted that after a change has been detected, normalisation  of the learning rate is carried out so as to ensure that the proposed method is independent of the signal scale. Finally, it should be noted that the proposed method assumes that the time index  differences between  change point locations are greater than $T_{r}$ samples, owing to the time period required for the estimation of the variance as well as the convergence of the filter to the `slow' volatility filter after a change has been detected. 

\indent Algorithm 2 proposes  the multivariate change detection algorithm via an extension of Algorithm 1. That is given a multichannel signal we seek to propose  a cooperative strategy between the data channels for improved detection of transition points.  This is achieved by utilising the combine then adapt (CTA) strategy of the diffusion LMS which was proposed in \cite{Sayed08};  where given $N_{sen}$ sensors such that each sensor $c$ has an input signal $\mathbf{x}_{c}(t)$, desired signal $d_{c}(t)$ and parameters $\mathbf{v}_{c}(t)$, where each sensor also has the ability to send and receive state estimates to one another\footnote{The diffusion LMS algorithm considers sensor networks, that is, all the sensors are not fully connected. However,  we consider devices where the same information is being recorded and thus we use full cooperation between all the sensors, in this work.}, accordingly the CTA strategy for cooperative adaptive filtering is given by \cite{Sayed08} 
\begin{equation}
\begin{aligned}
&\boldsymbol{\psi}_{c}(t)=\sum_{l\in \mathcal{N}_{c}}g_{c,l}\mathbf{v}_{l}(t-1)\\
&\mathbf{v}_{c}(t)=\boldsymbol{\psi}_{c}(t)+\mu_{c}\big(d_{c}-\boldsymbol{\psi}^{T}_{c}(t)\mathbf{x}_{c}(t)\big)\mathbf{x}_{c}(t)\\
\end{aligned}
 \label{HypothesisVolatility}
\end{equation}
 where $\mu_{c}$ is the learning rate of each sensor. The  update of the LMS is carried out with respect to the averaged estimates of the LMS output from the previous time instants, where $\mathcal{N}_{c}$ corresponds to the set of sensors connected to the $c^{th}$ sensor and $g_{c,l}$ corresponds to the combiner of the sensor estimates.  Algorithm 2 provides an overview of the cooperative adaptive filter based change detection (CAFCD) algorithm  for multivariate data. 
%\begin{equation}
%\sigma_{o}(t+1) = \lambda(t+1)\sigma_{f}(t+1) + (1-\lambda(t+1))\sigma_{s}(t+1).
%\end{equation}
\begin{algorithm}[t!]
%\textbf{Initialise:}
 %  \indent Set $s(1)=1$, significance level $k$ and volatility filter window size $T$.
\begin{algorithmic} 
\REQUIRE 
\STATE - Input signal $x_{c}(t)$ 
\STATE - Select: $\mu_{c}, T_{s,c}, T_{f,c},T_{d,c},\gamma_{c}$.
\STATE - Set: $\lambda_{c}(1)\leftarrow 1$, $t\leftarrow1$, $\rho_{c}\leftarrow0.001$, $T_{r,c}\leftarrow1.2T_{s,c}$.  \STATE $\psi(1)\leftarrow\sum_{c=1}^{N_{c}}\lambda_{c}(1)$
\vspace{1mm}
\WHILE{$t< N$}
\FOR{$c=\{1,\dots,N_{sen}\}$} 
\STATE -  $\hat{\sigma}_{o,c}(t) \leftarrow \psi(t)\hat{\sigma}_{f,c}(t) + (1-\psi_{s}(t))\hat{\sigma}_{s,c}(t)$.
\STATE -  $e_{c}(t)\leftarrow \hat{\sigma}_{d,c}(t)-\hat{\sigma}_{o,c}(t)$.
\STATE  - $\lambda_{c}(t+1) \leftarrow \psi_{c}(t) + \mu_{c}(|\psi(t)|+\rho_{c}u_{c})e_{c}(t)(\hat{\sigma}_{f,c}(t) - \hat{\sigma}_{s,c}(t))$
\ENDFOR
\STATE $\psi(t+1)\leftarrow\sum_{c=1}^{N_{c}}\lambda_{c}(t+1)$
\IF{$\psi(t+1)\geq\gamma_{c}$ and $\text{Flag}(t)=0$}
\STATE - $s(t)\leftarrow1$ in the interval $[t,t+T_{r}]$.
\STATE - $\mu_{c}\leftarrow \frac{\mu_{c}}{\text{Var}(x_{c}(t))}$ for $[t,t+T_{r}]$.
\STATE - $ \text{Flag}(t)=1$ in the interval $[t,t+T_{r}]$.
\ENDIF
\STATE - $t \leftarrow t+1$
\ENDWHILE
\end{algorithmic}
%\textbf{Output} 
\caption{\!\!: Cooperative Adaptive Filtering Based Change Detection}\label{alg:CooperativeAdaptiveFilteringBasedChange}
\end{algorithm}

\subsection{Volatility Filter Weight Selection}
In this section we  analyse the most appropriate choice of weights for both the `fast' $\boldsymbol{\text{w}}_{f}$ and `slow' $\boldsymbol{\text{w}}_{s}$ volatility filters, for detecting changes in time series data. Namely, given a sequence of zero mean data points, $x_{1},...x_{N}$ with a transition time at $t_{1}$, where the variance of the first segment is given by $\sigma^{2}_{1}$ and the variance of the second segment is given by $\sigma^{2}_{2}$. The objective is to determine, how the choice of volatility filter weights affects the detection performance of the adaptive filtering based change detection algorithm. 
\\
\indent Note that the detection performance is dependent on the rate at which $\lambda(t)\rightarrow 1$ after the transition time $t_{1}$. For rigour,  we calculate the minimum of the expectation of the cost function in \eqref{CF} to analyse the evolution of $\lambda(t)$. That is, we determine the expected value of $\lambda(t)$ that minimises the cost function  at each time index, shown by the following
\begin{equation}
\begin{aligned}
\text{E}\{\lambda(t)\} &= \frac{\text{E}\left\{(\hat{\sigma}_{d}(t)-\hat{\sigma}_{s}(t))(\hat{\sigma}_{f}(t)-\hat{\sigma}_{s}(t))\right\}}{\text{E}\left\{(\hat{\sigma}_{f}(t)-\hat{\sigma}_{s}(t))^{2}\right\}}
\end{aligned}
 \label{minlambda}
\end{equation}
% \begin{equation}
% \begin{aligned}
% \text{E}\{\lambda(t)\} &= \text{E}\left\{\frac{(\hat{\sigma}_{d}(t)-\hat{\sigma}_{s}(t))(\hat{\sigma}_{f}(t)-\hat{\sigma}_{s}(t))}{(\hat{\sigma}_{f}(t)-\hat{\sigma}_{s}(t))^{2}}\right\}\\\\
% &\approx \frac{\text{E}\left\{(\hat{\sigma}_{d}(t)-\hat{\sigma}_{s}(t))(\hat{\sigma}_{f}(t)-\hat{\sigma}_{s}(t))\right\}}{\text{E}\left\{(\hat{\sigma}_{f}(t)-\hat{\sigma}_{s}(t))^{2}\right\}}
% \end{aligned}
%  \label{minlambda}
% \end{equation}

\begin{figure}[t!]
\begin{center}
\includegraphics[width=0.5\columnwidth]{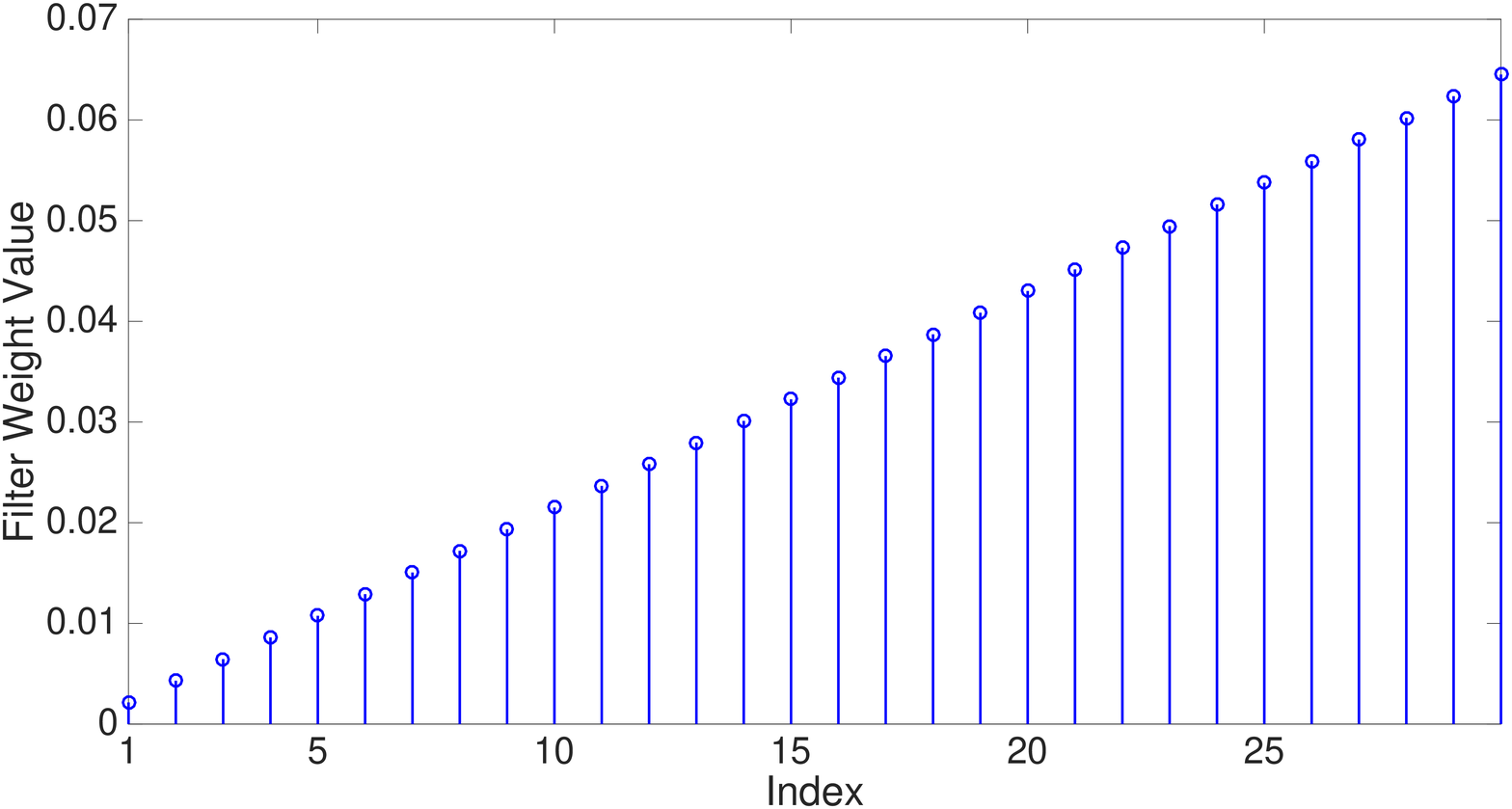}
\includegraphics[width=0.5\columnwidth]{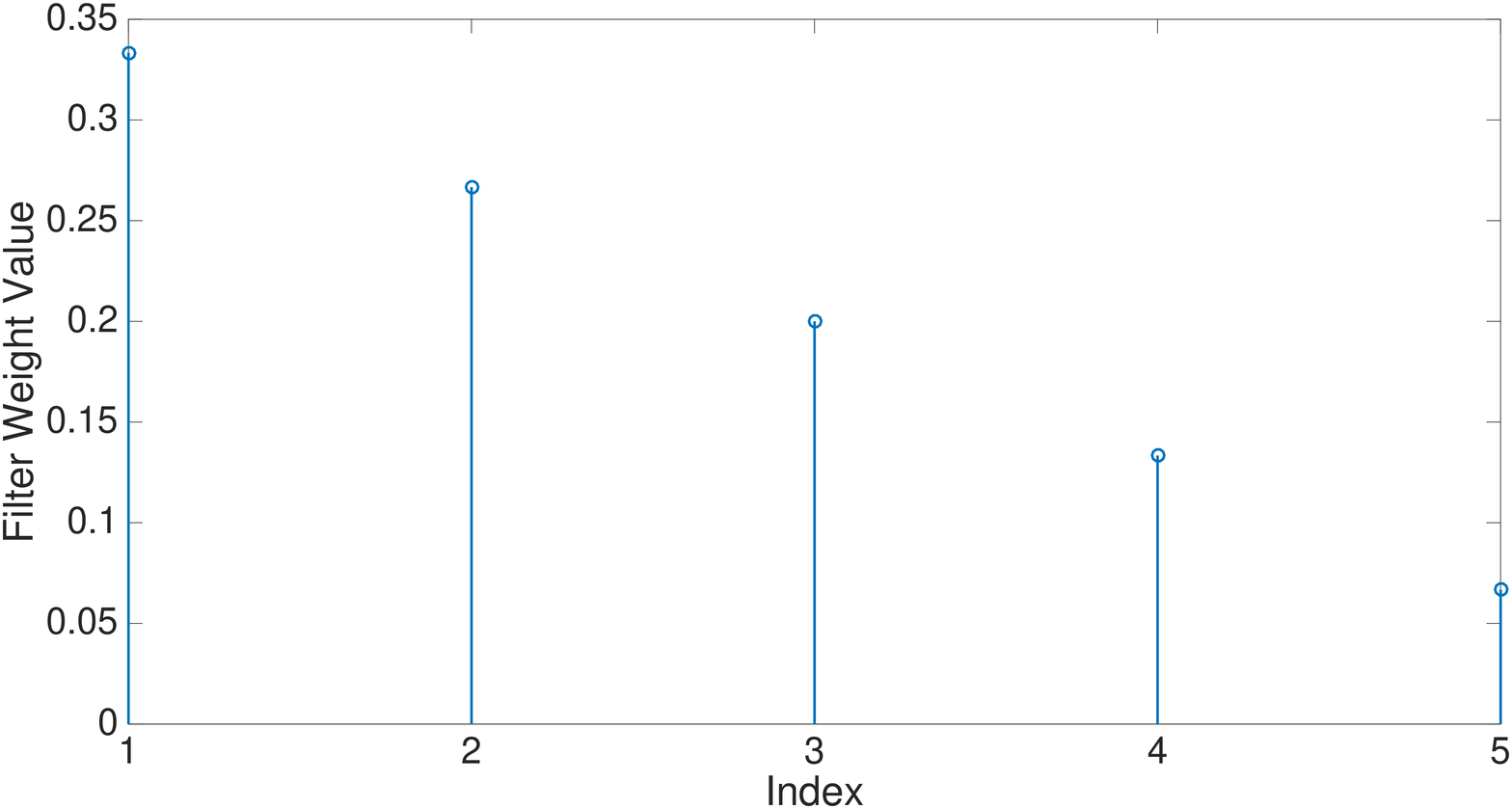}
\caption{Figure showing the triangular impulse response function for  `slow' filter (upper panel) and  `fast'  filter (lower panel).  
 }\label{fig:FiltExample}
\end{center}
\end{figure}

Given a change point at time $t_{1}$, the expression in \eqref{minlambda} can be evaluated at three different time intervals: 1) $t<t_{1}$ and $t>t_{1}+T_{s}$, 2) $t_{1}\leq t \leq t_{1}+T_{f}$, and 3) $t_{1}+T_{f}<t\leq t_{1}+T_{s}$. 
% \indent Before proceeding to analyse the behaviour for the estimate of $lambda(t)$, for a given set of volatility filter weights, we will first verify The set of time intervals in 1) correspond to the case where both the fast `fast' and `slow' volatility filters are measuring data points such that statistics of the data points are identical (stationary). The expression in \eqref{minlambda} can be evaluated by substituting the following values: $\text{E}\{\hat{\sigma}_{d}\}=\text{E}\{\hat{\sigma}_{s}\}=\text{E}\{\hat{\sigma}_{f}\}\approx \sigma_{i}$, $\text{E}\{\hat{\sigma}^{2}_{s}\}=\text{E}\{\hat{\sigma}^{2}_{f}\}=\sigma^{2}_{i}$ and the covariance $\text{Cov}(\hat{\sigma}_{s},\hat{\sigma}_{f})=\epsilon$, for $i=1,2$. This results in $\text{E}\{\lambda_{\text{min}}(t)\} \approx \frac{1}{2}$.
\\
\indent Now the set of time intervals in 1) correspond to the case where both the  `fast' and `slow' volatility filters are measuring data points such that statistics of the data points are identical (strictly stationary). It is straightforward to show that $\text{E}\{\lambda(t)\} \approx \frac{1}{2}<1$.  The time intervals in 2) and 3) correspond to the behaviour of $\lambda(t)$ after a change has occurred. In order to evaluate the performance of different window parameters on the behaviour of $\lambda(t)$, we compute \eqref{minlambda} over the interval\footnote{A similar argument can be made for interval 3).} 2). Moreover, the interval  2) corresponds to the case where the input data points to both the `fast' and `slow' volatility filters are equivalent to a mixture of both the first segments statistics with expected variance $\sigma^{2}_{1}$ and the second segments statistics with expected variance $\sigma^{2}_{2}$. That is, the expected value of the output for the squared volatility filter with weights $\boldsymbol{\text{w}_{f}}$, is given by 
\begin{equation}
\text{E}\{\hat{\sigma}^{2}_{f}(t)\}= w^{1}_{f}(t)\sigma_{1}^{2}+w^{2}_{f}(t)\sigma^{2}_{2}
 \label{sigma_f}
\end{equation}
with $\text{E}\{\hat{\sigma}_{f}(t)\}\approx \sqrt{\text{E}\{\hat{\sigma}^{2}_{f}(t)\}}$,   for $t_{1}\leq t \leq t_{1}+T_{f}$, where $w^{1}_{f}(t)$ is a monotonically increasing function  and $w^{2}_{f}(t)$ monotonically decreasing function, such that both functions are dependent on the volatility filter  weights $\boldsymbol{\text{w}_{f}}$, with sum of the two functions  equal to one. The corresponding output of the squared volatility filter with weights $\boldsymbol{\text{w}_{s}}$, is given by 
\begin{equation}
\text{E}\{\hat{\sigma}^{2}_{s}(t)\}=w^{1}_{s}(t)\sigma_{1}^{2}+w^{2}_{s}(t)\sigma^{2}_{2}
 \label{sigma_s}
\end{equation}
with $\text{E}\{\hat{\sigma}_{s}(t)\}\approx \sqrt{\text{E}\{\hat{\sigma}^{2}_{s}(t)\}}$, for $t_{1}\leq t \leq t_{1}+T_{s}$,  where $w^{1}_{s}(t)$  and $w^{2}_{s}(t)$ are monotonically increasing and decreasing functions respectively, with the sum of both functions  equal to one. Accordingly, \eqref{minlambda} evaluated over  $t_{1} \leq t \leq t_{1}+T_{f}$, is given by
\begin{equation}
\text{E}\{\lambda(t)\}\approx \frac{A(t)}{B(t)}
 \label{expLambda}
\end{equation}
where
\begin{equation}
\begin{aligned}
A(t)&\approx \sigma_{2}\big(\text{E}\{\hat{\sigma}_{f}(t)\}-\text{E}\{\hat{\sigma}_{s}(t)\}\big)-\text{Cov}\big(\hat{\sigma}^{2}_{s}(t),\hat{\sigma}^{2}_{f}(t)\big)\\
&+\text{E}\{\hat{\sigma}^{2}_{s}(t)\}-\text{E}\{\hat{\sigma}_{f}(t)\}\text{E}\{\hat{\sigma}_{s}(t)\}
 \label{A}
 \end{aligned}
\end{equation}
and
\begin{equation}
\begin{aligned}
B(t)&\approx \text{E}\{\hat{\sigma}^{2}_{f}(t)\}+\text{E}\{\hat{\sigma}^{2}_{s}(t)\}-2\text{Cov}\big(\hat{\sigma}^{2}_{s}(t),\hat{\sigma}^{2}_{f}(t)\big)\\
&-2\text{E}\{\hat{\sigma}_{f}(t)\}\text{E}\{\hat{\sigma}_{s}(t)\}
 \label{B}
 \end{aligned}
\end{equation}
where $\text{E}\{\hat{\sigma}_{d}(t)\}\approx\sigma_{2}$ and $\text{Cov}\big(\hat{\sigma}^{2}_{s}(t),\hat{\sigma}^{2}_{f}(t)\big)=\epsilon$, where $\epsilon\rightarrow0$ for $T_{s}>>T_{f}$. From  \eqref{sigma_f}, \eqref{sigma_s} and \eqref{expLambda} for $\lambda(t)\rightarrow 1$, we require $\text{E}\{\hat{\sigma}^{2}_{f}(t)\}\approx \sigma_{1}$ and $\text{E}\{\hat{\sigma}^{2}_{s}(t)\}\approx \sigma_{2}$ over the interval $t_{1} \leq t \leq t_{1}+T_{f}$. The analysis implies that for the `fast' volatility filter a higher weighting needs to be applied to the most recent samples in $\boldsymbol{\text{w}}_{f}$. While for the `slow' volatility filter  the converse is true, lower weighting needs to be applied to the most recent samples in $\boldsymbol{\text{w}}_{s}$. An example of such a pair of  filters with impulse responses that satisfy the condition above are  shown in Fig.~\ref{fig:FiltExample}, namely a pair of triangular weighted filters, that is
\begin{equation}
 \hat{\sigma}^{2}_{s}(t) = \frac{x^{2}_{t}+2x^{2}_{t-1}+...+(T_{s}-1)x^{2}_{t-T_{s}+2}+T_{s}x^{2}_{t-T_{s}+1}}{1+2+...+T_{s}} \label{tri_slow}
\end{equation}
for the 'slow' volatility filter weights, and 
\begin{equation}
\hat{\sigma}^{2}_{f}(t) = \frac{T_{f}x^{2}_{t}+(T_{f}-1)x^{2}_{t-1}+...+2x^{2}_{t-T_{f}+2}+x^{2}_{t-T_{f}+1}}{1+2+...+T_{f}} \label{tri_fast}
\end{equation}
for the 'fast' volatility filter weights.

\vspace{-3mm}
\subsection{Volatility Change Point Estimator}
\begin{figure}[t!]
\begin{center}
\includegraphics[width=0.5\columnwidth]{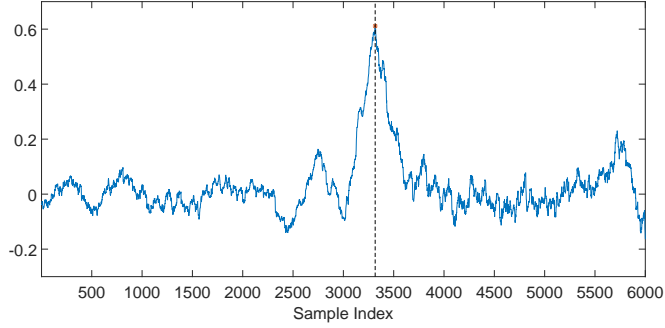}
\caption{The output of the volatility filter (of window size $T$)  (dashed line).
 }\label{fig:ChangeSynthetic}
\end{center}
\end{figure}

The second method referred to as the volatility change estimator (VCE), seeks to estimate
the location of the change point transition times. This is achieved by first
obtaining the differenced\footnote{That is the difference between the output of the
 volatility filter at the current time index $t$  and  the output
 of the volatility filter at time index $t-T_{l}$.} estimate $\hat{\sigma}_{D}(t)$ from the output of the volatility filter
$\hat{\sigma}_{l}(t)$, as follows
\begin{equation}
\hat{\sigma}_{D}(t) = \hat{\sigma}_{l}(t)-\hat{\sigma}_{l}(t-T_{l}), \label{difference}
\end{equation}
where   $T_{l}$ corresponds to the window size and $ \hat{\sigma}_{l}(t)$ is estimated using the square window  volatility filter weights $\boldsymbol{\text{w}}_{l}=\frac{1}{T_{l}-1}\boldsymbol{1}_{T_{l}}$ (where $\boldsymbol{1}_{T_{l}}$ is a vector of length $T_{l}$ with each element equal assigned the value $1$).
 In order to develop a change point
location estimator, we first observe the following property in
the plot of $\hat{\sigma}_{D}(t)$ with respect to sample indices. Consider  a zero mean piecewise stationary (with respect to variance) Gaussian distributed sequence of data points  with a change in the variance at the time index 3001. The corresponding plot of  $\hat{\sigma}_{D}(t)$ is shown in Fig.~\ref{fig:ChangeSynthetic}. Observe
that $\hat{\sigma}_{D}(t)$ oscillates (with an approximately constant variation) around the zero until the change point at sample index $3001$, where $\hat{\sigma}_{D}(t)$ increases significantly (with respect to the prior variation in $\hat{\sigma}_{D}(t)$) and at approximately $3000+T_{l}$ samples the time series hits a maximum (shown by the dashed line in Fig.~\ref{fig:ChangeSynthetic}), as a result the following estimator is proposed as a change point location estimate
\begin{equation}
    \hat{\tau} =  \argminA_t|\hat{\sigma}_{D}(t)| - T_{l}+1. \label{estimator}
\end{equation}
While the proposed change point location estimate can be shown to be a reasonable unbiased estimator for a single volatility change in the time series data (see the following section), a set of rules are required to deal with multiple change point locations. That is, if a change has
been detected by the AFCD algorithm, the estimator in \eqref{estimator}
is applied in the interval between the change point detection
time to $2T_{l}$ samples after the detection. For multivariate data, either an estimate of the change point location for each channel is  obtained, or a single estimate of the change point location is obtained by averaging across the individual channel estimates.
\vspace{-3mm}
 \subsection{Change Point Location Estimator Analysis}
 In this section we verify that the proposed change point estimator \eqref{estimator} is a reasonable unbiased estimator for the  transition point location. Given a sequence of zero mean Gaussian random variables, 
$x_{1},\dots,x_{\tau-1}\in N(0,\sigma^{2}_{1})$ and $x_{\tau},\dots,x_{N}\in N(0,\sigma^{2}_{2})$, where the change point occurs at the time index $\tau$,  the following equality for \eqref{estimator} is satisfied
\begin{equation}
   \hat{\tau}  = \text{E}\left\{\argminA_{t}|\hat{\sigma}_{D}(t)| \right\}- T_{l}+1 \label{Eestimator}
\end{equation}
in which $T_{l}$ is the volatility filter window size and $\argminA_{t}|\hat{\sigma}_{D}(t)|$ is equivalent to showing the following statement
 \begin{equation}
\begin{aligned}
\text{E}\{|\hat{\sigma}_{D}(\tau+T_{l}-1)|\}>\text{E}\{|\hat{\sigma}_{D}(\tau+i)|\} \quad i\neq T_{l}-1
\end{aligned}
 \label{ExpectationCondition}
\end{equation}
where $i$ is the set of time shifts such that there is no further transitions in volatility  within $i<<N$ samples of  $\tau$. This condition ensures that the maximum evaluated for $\argminA_{t}|\hat{\sigma}_{D}(t)|$, is related to a single instantaneous volatility change. For multiple instantaneous transitions in volatility, the spacing between any two such transition time points are assumed to be greater than $i>2T_{l}$. If the following volatility change conditions $\sigma_{2}\gtrless \sigma_{1}$ hold, then \eqref{ExpectationCondition} is equivalent to 
 \begin{equation}
\begin{aligned}
\text{E}\{\hat{\sigma}_{D}(\tau+T_{l}-1)\}\gtrless \text{E}\{\hat{\sigma}_{D}(\tau+i)\} \quad i\neq T-1
\end{aligned}
 \label{ExpectationCondition1}
\end{equation}
For  $\sigma_{2}< \sigma_{1}$ (where an equivalent argument can be made for  $\sigma_{2}> \sigma_{1}$), 
 \begin{equation*}
\begin{aligned}
\hspace{-3mm}\text{E}\{\hat{\sigma}_{D}(\tau+T_{l}-1)\}&=\text{E}\{\hat{\sigma}_{l}(\tau+T_{l}-1)\}-\text{E}\{\hat{\sigma}_{l}(\tau-1)\}\\
\hspace{-6mm}&=\hspace{-1mm}\text{E}\hspace{-0.5mm}\left\{\hspace{-1.5mm}\sqrt{\sum_{j=0}^{T_{l}-1} \frac{x^{2}_{\tau+T_{l}-1-j}}{T_{l}-1}}\right\}\hspace{-1mm}-\hspace{-1mm}\text{E}\hspace{-0.5mm}\left\{\hspace{-1.5mm}\sqrt{\sum_{j=0}^{T_{l}-1} \frac{x^{2}_{\tau-1-j}}{T_{l}-1}}\right\}
\end{aligned}
 \label{Expectation1}
\end{equation*}
we next take the first order Taylor expansion 
 \begin{equation}
\begin{aligned}
\text{E}\{\hat{\sigma}_{D}(\tau+T_{l}-1)\}&\approx\text{E}\hspace{-0.5mm}\left\{\sum_{j=0}^{T_{l}-1} \frac{x^{2}_{\tau+T-1-j}}{T_{l}-1}\right\}^{\frac{1}{2}}\hspace{-3mm}-\hspace{-1mm}\text{E}\hspace{-0.5mm}\left\{\sum_{j=0}^{T_{l}-1}\frac{ x^{2}_{\tau-1-j}}{T_{l}-1}\right\}^{\frac{1}{2}}
\end{aligned}
 \label{Expectation2}
\end{equation}
where it should be noted that the sum of squared  standard normal distributed variables  are from a chi-squared distribution, that is,
 \begin{equation*}
\begin{aligned}
&\sum_{j=0}^{T_{l}-1} \frac{x^{2}_{\tau+T_{l}-j}}{\sigma_{2}^{2}}\backsim \chi^2(T_{l}-1)\\
&\sum_{j=0}^{T_{l}-1} \frac{x^{2}_{\tau-j}}{\sigma_{1}^{2}}\backsim \chi^2(T_{l}-1)\\
\end{aligned}
 \label{Distribtuion}
\end{equation*}
As a result, $\text{E}\{\hat{\sigma}_{D}(\tau+T_{l}-1)\}\approx \sigma_{2}-\sigma_{1}$. The corresponding estimates for $\text{E}\{\hat{\sigma}_{D}(t+T_{l}+j)\}$ and $\text{E}\{\hat{\sigma}_{D}(t+T_{l}-j)\}$ for $k=2,...,T_{l}$ are given by 
 \begin{equation*}
\begin{aligned}
\text{E}\{\hat{\sigma}_{D}(\tau+T_{l}+k)\}\approx \sigma_{2}-\sqrt{\frac{(T_{l}-k)\sigma^{2}_{1} +(k-1)\sigma_{2}^{2}}{T_{l}-1}}
\end{aligned}
 \label{Distribtuion}
\end{equation*}
 \begin{equation*}
\begin{aligned}
\text{E}\{\hat{\sigma}_{D}(\tau+T_{l}-k)\}\approx \sqrt{\frac{(T_{l}-k)\sigma^{2}_{2} +(k-1)\sigma_{1}^{2}}{T_{l}-1}}-\sigma_{1}
\end{aligned}
 \label{Distribtuion}
\end{equation*}
where it can be  shown  that $\text{E}\{\hat{\sigma}_{D}(\tau+T_{l}-1)\}$ is less than both $\text{E}\{\hat{\sigma}_{D}(\tau+T_{l}+k)\}$ and $\text{E}\{\hat{\sigma}_{D}(\tau+T_{l}-k)\}$, thus illustrating that, $\argminA_{t}|\hat{\sigma}_{D}(t)|$ provides a reasonable unbiased estimate of the change point location. 
\begin{figure}[t!]
\begin{center}
\includegraphics[width=0.5\columnwidth]{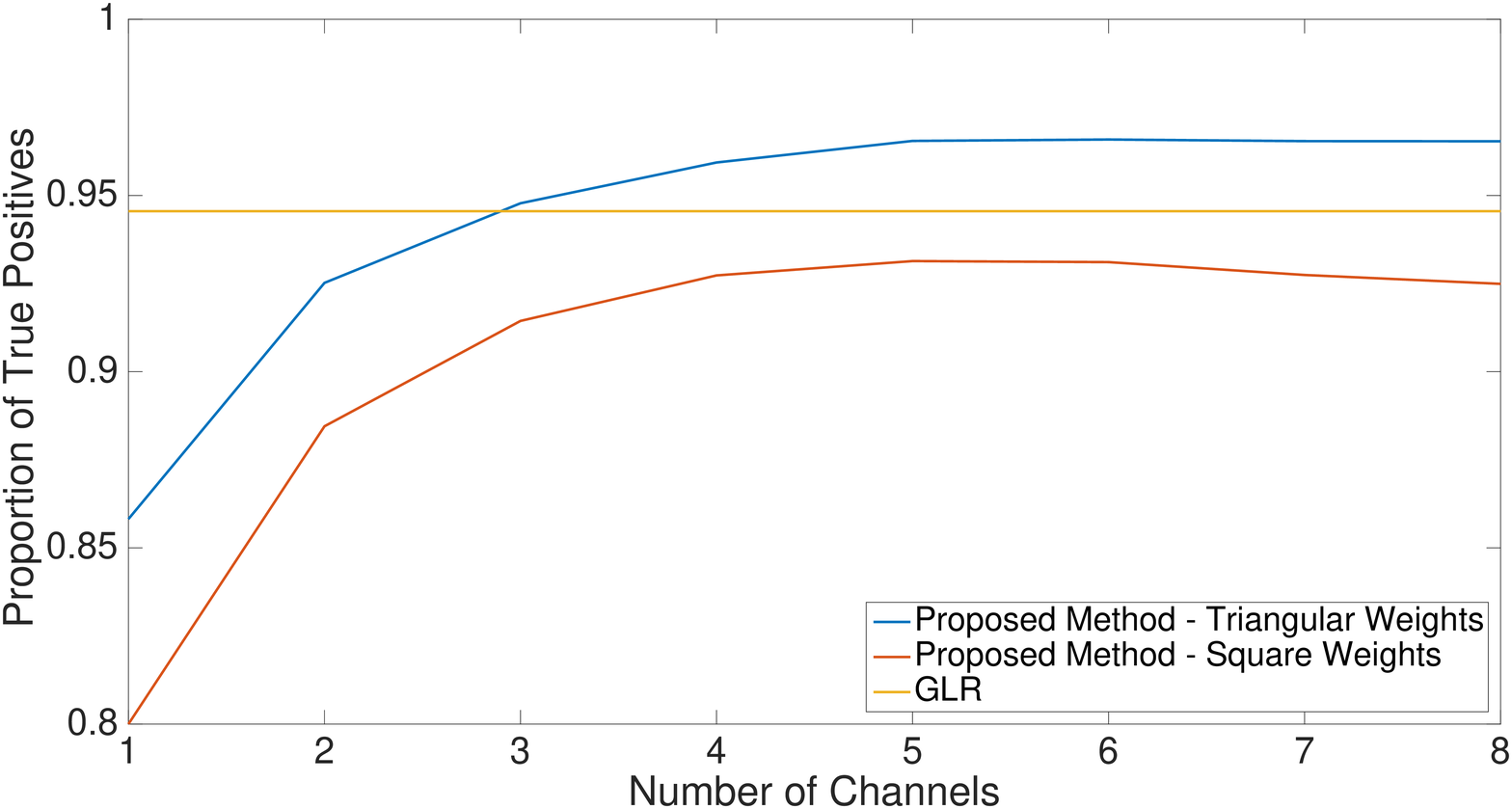}
\includegraphics[width=0.5\columnwidth]{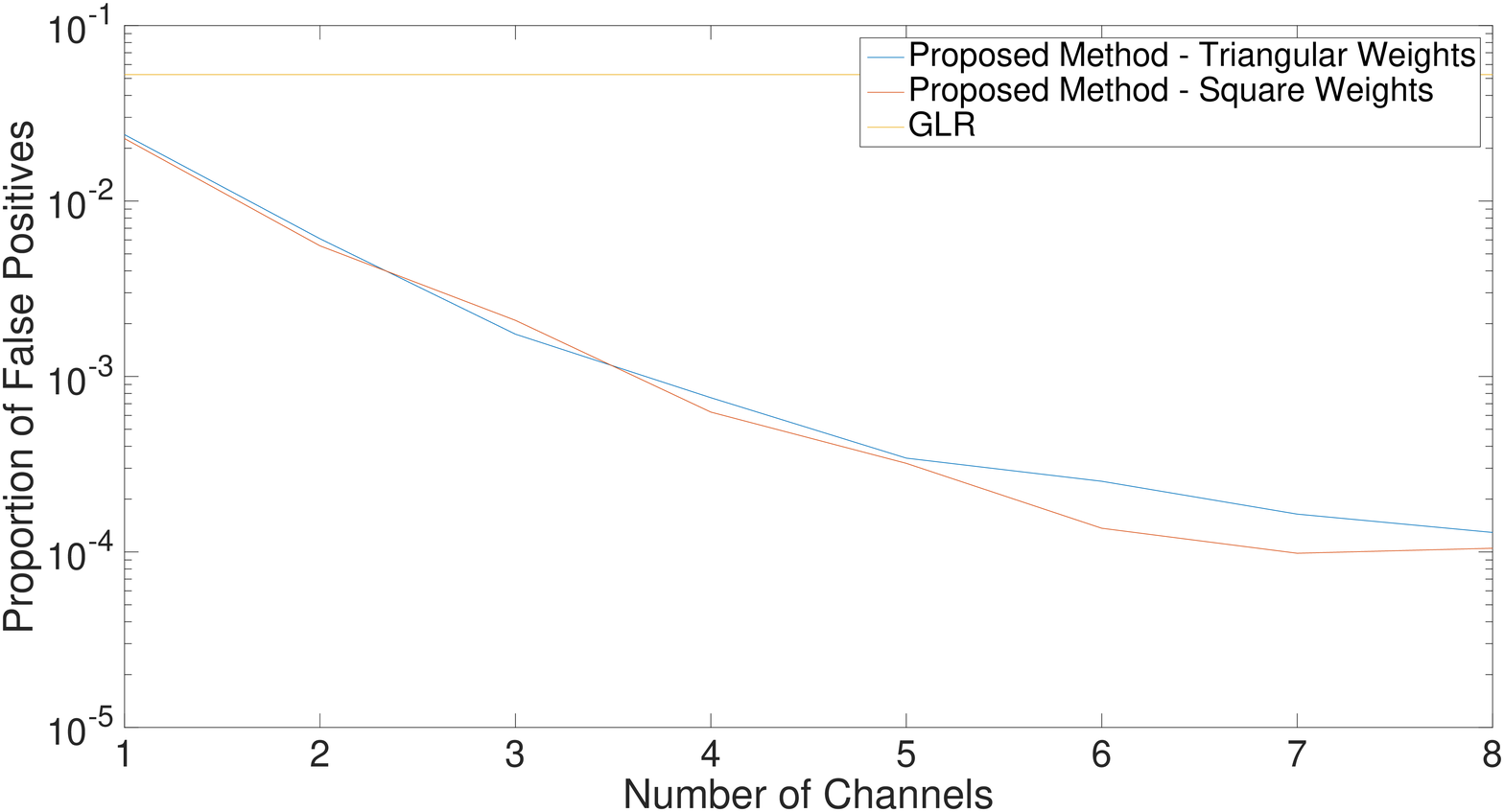}
\includegraphics[width=0.5\columnwidth]{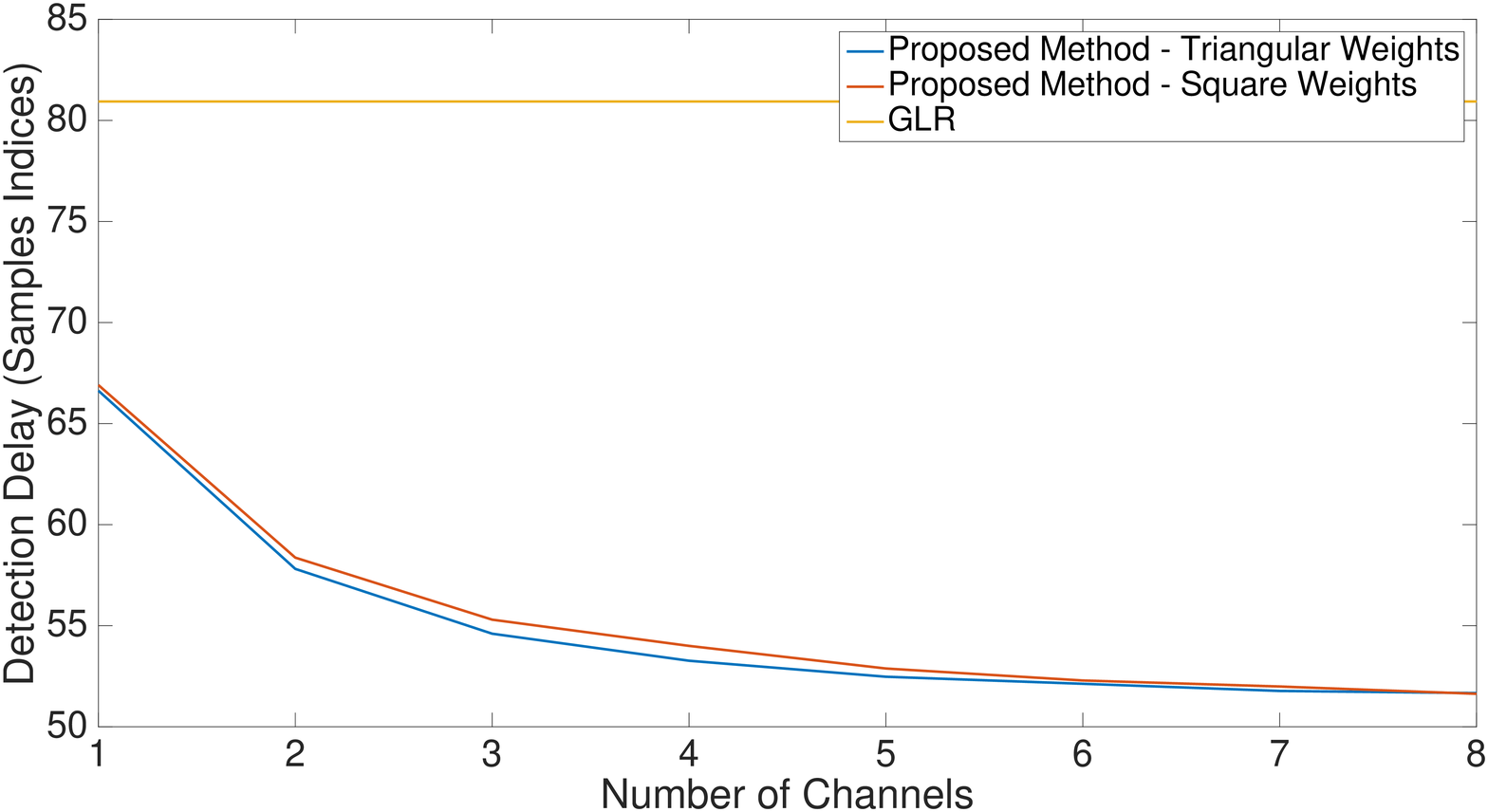}
\caption{Figures comparing the proposed method (for both triangular and square filter weights) along with the GLR. (Upper panel) Proportion of true positives, (middle panel) proportion of false positives and (lower panel) detection latency. Multivariate data generated using random correlation matrices with low correlations. 
 }\label{fig:SimulationCorr_0}
\end{center}
\end{figure}

\begin{figure}[t!]
\begin{center}
\includegraphics[width=0.5\columnwidth]{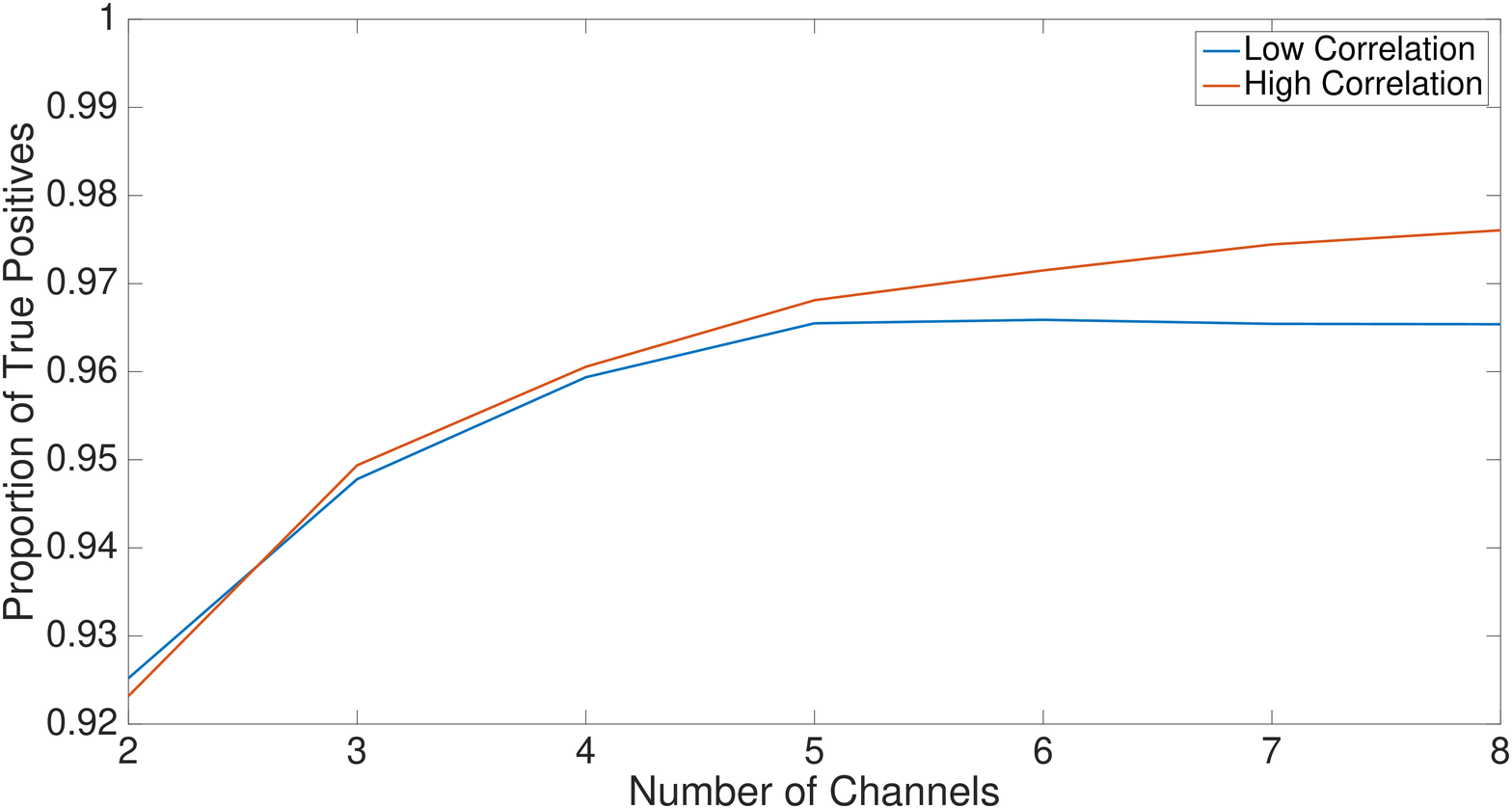}
\includegraphics[width=0.5\columnwidth]{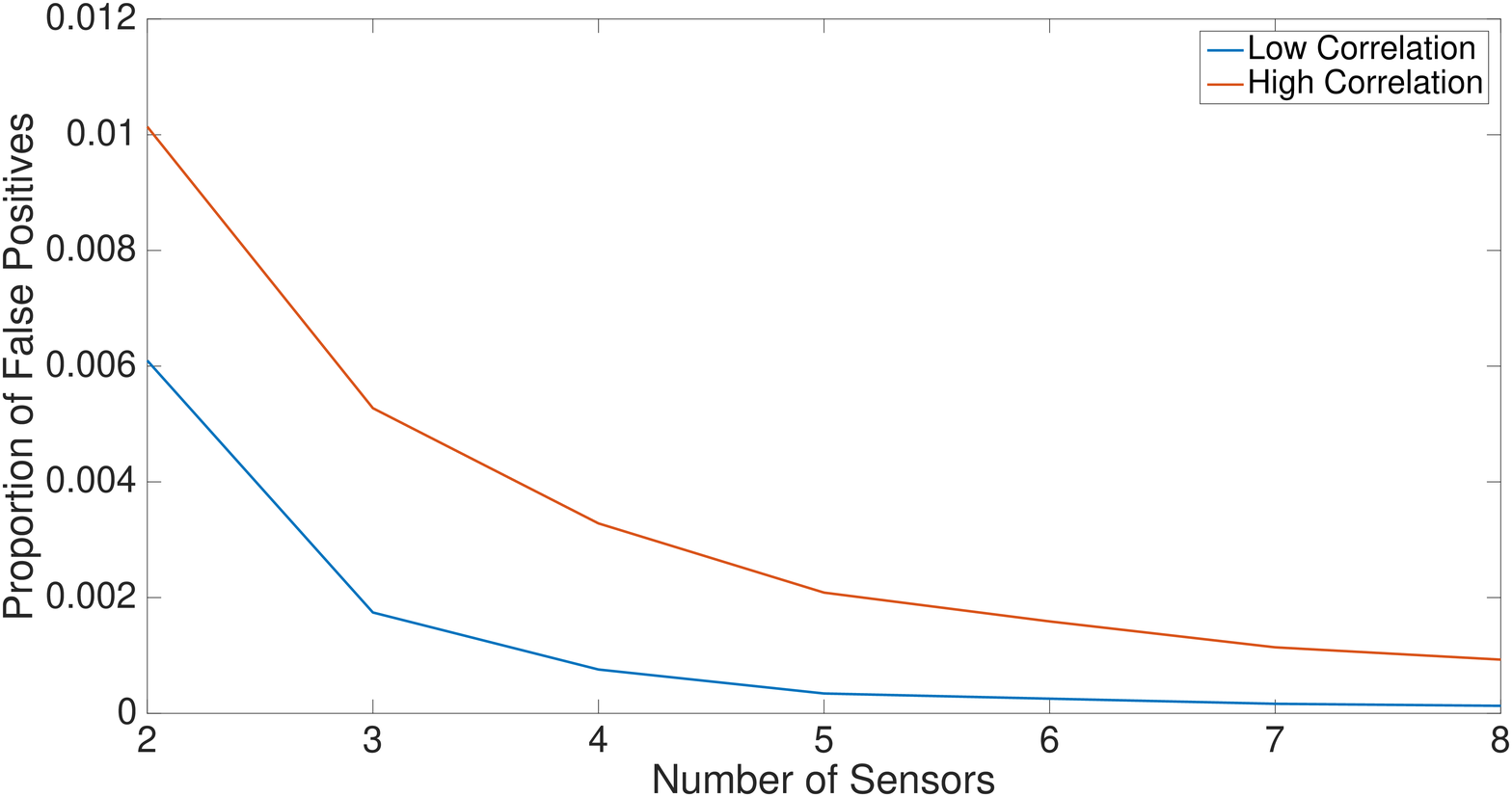}
\includegraphics[width=0.5\columnwidth]{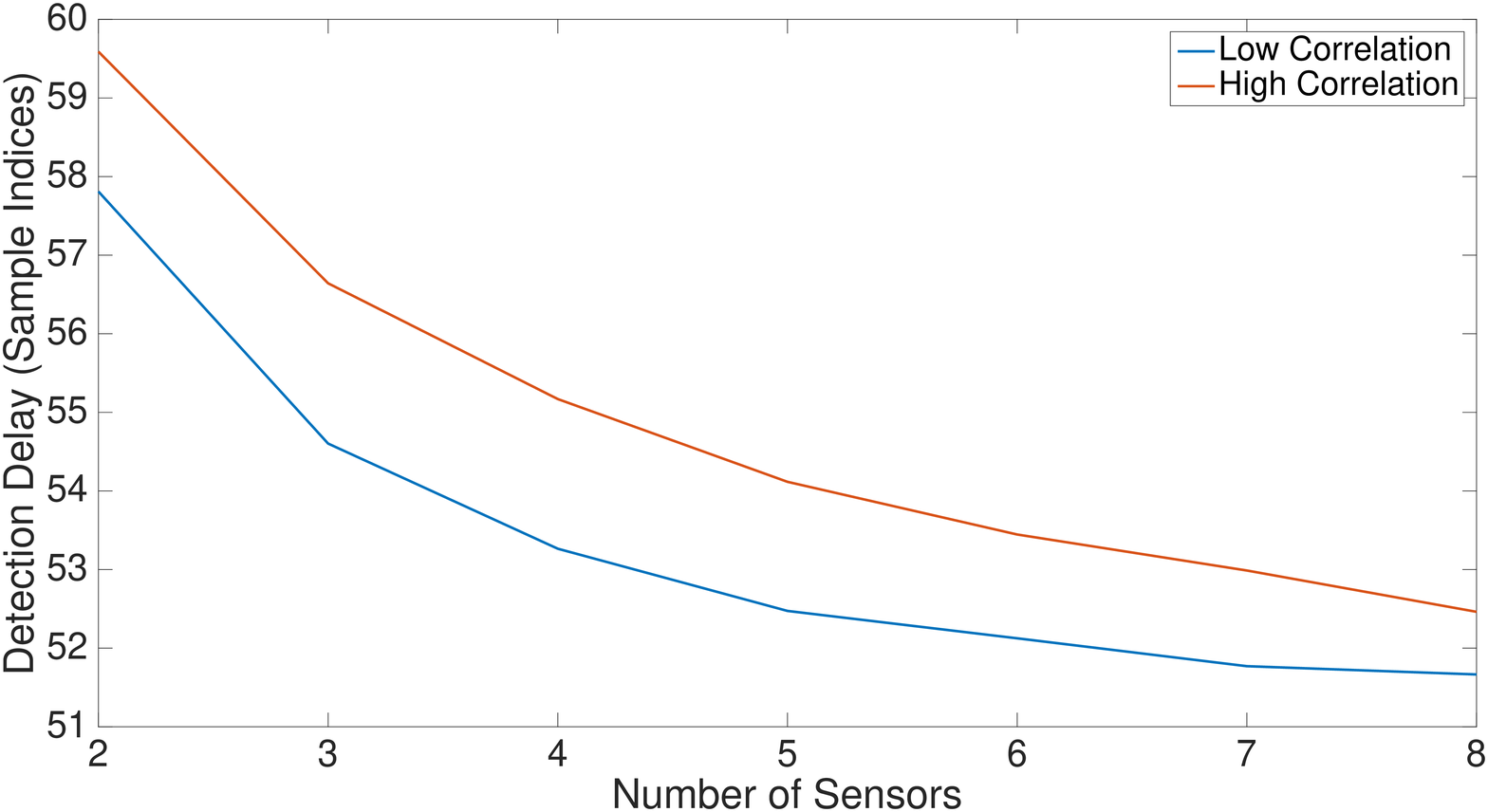}
\caption{Figures comparing the proposed method with triangular  filter weights with multivariate data generated using random correlation matrices with high and low correlations. (Upper panel) Proportion of true positives, (middle panel) proportion of false positives and (lower panel) detection latency. 
 }\label{fig:SimulationCorr}
\end{center}
\end{figure}
 \vspace{-1mm}
\section{Simulations}
The proposed  change point detection and location estimation methods were analysed on both synthetic and real world data. In particular, we assessed the performance of the proposed method with the generalized likelihood ratio (GLR) method \cite{Brandt83}, owing to the  similar computational costs of both algorithms. 

% \begin{figure*}[!h]
% \centering \subfigure{
% \includegraphics[width=1\columnwidth]{TrueNegatives.eps}
% \includegraphics[width=1\columnwidth]{TrueNegatives.eps}
% \label{fig:AMFM10}}
% %
% \hspace{-3cm} \subfigure{
% \includegraphics[width=1\columnwidth]{FalsePositives.eps}
% \includegraphics[width=1\columnwidth]{FalsePositives.eps}
% \label{fig:AMFM5} }
% %
% \hspace{-3cm} \subfigure{
% \includegraphics[width=1\columnwidth]{DetectionDelay.eps}
% \includegraphics[width=1\columnwidth]{DetectionDelay.eps}
% \label{fig:AMFM0} } \caption[Optional caption for list of
% figures]{ The time-frequency representations for both the proposed
% method (left panels) and the MWPD (right panels) for a bivariate
% AM/FM signal, with input SNR of (a)  10 dB, (b) 5 dB and (c) 0 dB.
% The window length used for  MPWD was  681
% samples.}\label{fig:SimulationCorr_0}
% \end{figure*}

\subsection{Synthetic Signal}
The first synthetic simulation consists of both   univariate/multivariate time series  data points $\mathbf{x}(n)$ drawn from a univariate/multivariate Gaussian distribution with instantaneous changes in the variance at random points in time; furthermore,  each realization of the multivariate data points had a randomly  generated  correlation matrix  \cite{Lewandowski09}. The instantaneous power changes for each segment were selected randomly along with the  number of samples in the segment as well as the total signal length. The total number of samples was selected uniformly between $[5000,30000]$ and each  segment length was selected with uniform probability between the interval $[300,700]$. Furthermore, the variance for each segment was dependant on the variance of the prior segment; that is, a scale increase or decrease was selected with equal probability. The magnitude of the scale decrease was selected with uniform probability between $[0.5,0.85]$ and similarly for a scale increase $[1.2,1.7]$.
\\
 \indent We  evaluate the performance of the proposed method when using different window parameters. Namely, we assess the performance of the triangular window weights (selected from the analysis in Section III.B) shown in Fig.~\ref{fig:FiltExample}, along with the standard square window weightings (that is a sliding window estimate of the standard deviation). In order to quantitatively evaluate the performance of the respective algorithms, we utilised the following measures: the number of false positives  and true positives, the detection latency (the time between the event occurring and the corresponding first detection) and the absolute error in the  change point location estimate. The following parameters were selected for the respective algorithms: 1) GLR: window size 250, threshold 5, 2) C/AFCD (for both the triangular and square filter parameters): $T_{s}=250$, $T_{f}=20$, $\gamma=0.8$, $T_{d}=10$, furthermore we selected the learning rate $\mu$, such that the proportion of false positives for both the triangular and square filter weights   were equal. 
\\
\indent Fig.~\ref{fig:SimulationCorr_0} shows the results of the  proposed method (with the two different window parameters) along with the GLR method when analysing univariate and multivariate data (where for this simulation we utilised randomly generated correlation matrices with low absolute value in the off-diagonal entries). Observe that the proposed method (for both the triangular and square window weights) outperformed the GLR with respect to the proportion of false positives and detection latency. Furthermore, it can be seen that the proportion of true positives for the  proposed method with  triangular filter weights outperformed the GLR when the number of sensors was greater than two; while the proposed method with square window weights performed poorly. Finally, the results indicate that the weighting parameter for the volatility filters identified in section III.B  outperformed the standard square window volatility filter weights, with respect to the proportion of true positives. 
\\
\begin{figure}[t!]
\begin{center}
\includegraphics[width=0.5\columnwidth]{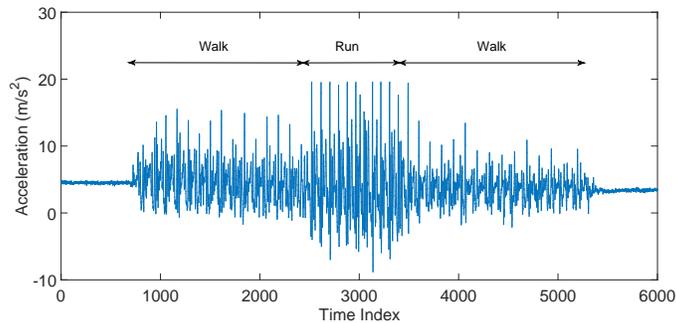}
\caption{Figure showing the time domain representation of the accelerometer data, where the three different states of motion in time have been highlighted. 
 }\label{fig:Accelerometer}
\end{center}
\end{figure}
\indent We also assessed the performance of the proposed method with randomly generated correlation matrices with high absolute values in the off-diagonal entries. In particular Fig.~\ref{fig:SimulationCorr} compares the randomly generated high  and low correlation matrices when using  volatility filters with triangular weights (developed in section III.B). Observe that the detection delay and proportion of false positives along with the true positives increase when there are large inter-channel correlations. This implies that for highly correlated data, the proposed method is more susceptible to detecting changes, owing to the proportion of false positives doubling, while the proportion of true positives does not increase significantly. 
\\
\indent Finally, we compared the performance of the proposed change point location estimator with the GLR's, where we evaluated the performance of the respective algorithms using synthetic univariate time series data. In particular, we found that the proposed VCE had an absolute error in the change point location estimate of $4.09$ samples, while the GLR's change point location estimator had an error of $30.34$ samples.

\begin{table}[t!]
\caption{ The performance of the CAFCD and GLR (applied to each channel) algorithms on real world accelerometer data.}
%\vspace{-4mm}
\begin{center}
\large\addtolength{\tabcolsep}{-5pt} { \small
\begin{tabular}{|c|c|c|c|}   \hline %\hline
 \begin{tabular}[x]{@{}c@{}}False Negative\\CAFCD\end{tabular}&~\begin{tabular}[x]{@{}c@{}}False Negative\\GLR\end{tabular}
&~\begin{tabular}[x]{@{}c@{}}Proportion\\CAFCD$<$GLR \end{tabular}&~\begin{tabular}[x]{@{}c@{}}Latency\\ (No. of Samples)\end{tabular}
\\ \hline
$0.927$  & $0.8152$ & $ 0.64$ & $14.78$ \\
 \hline
\end{tabular}}
\end{center}\label{SimChange}
\label{Tab:SimChange}
\end{table}%

\subsection{Accelerometer Data}
The final real world simulation assessed the performance of the proposed algorithm on real world tri-axial accelerometer data obtained from a  smart phone attached to  test subjects (where the data was obtained from \cite{Brajdic13} that consisted of 23 experimental trials with sampling frequency of 128 Hz).  Each experimental trial consists of five different states of motion: `no motion', `walking', `running', `walking' and `no  motion'; where an example from the output of  the accelerometer is shown in Fig.~\ref{fig:Accelerometer}. First differencing of the data was first carried out in order to remove trend components in the data.  We then compared the proposed CAFCD detection method (using the triangular filter weights scheme outlined in Section III.B) with the  GLR method that was applied seperately to each data channel.  While the underlying  true change point locations are unknown, the four different changes in the volatility of the accelerometer data can be observed via visual inspection (see Fig.~\ref{fig:Accelerometer}). As a result, we can obtain the following quantitative performance measures: proportion of true positives, the proportion of the CAFCD algorithm detecting the change in variance before GLR and the relative  difference between the detection latencies of GLR and CAFCD (that is a positive number would indicate CAFCD of having a smaller detection latency from the true change point location). From Table~\ref{Tab:SimChange}, it can be observed that the proposed change detection method had a lower number of false negatives when compared with the GLR algorithm (applied to each channel). Furthermore, the proposed method was able to detect more rapidly the transitions in 64\% percent of the changes in the states of motion. 
\section{Conclusion}
This work developed a class of sequential volatility filtering based change detection  and location estimation algorithms for time series data with instantaneous changes  in the volatility. The first method developed a change point detection  method based on adaptive filtering, and a convex combination of two volatility filters with differing window sizes.  The adaptive filtering formulation of the change point problem also enabled a straightforward extension to the analysis of changes in volatility in  multivariate time series data. This work also developed a change point location estimator based on the differenced output of a volatility filter.  The performance of the proposed detection and location estimation methods were evaluated on both synthetic and real world data sets.   

%\section{Code Availability}
%The code for this work has been made available at\footnote{The link for this code has been removed as per the guidelines for the peer review process of IEEE transactions on industrial informatics.}. 

\bibliographystyle{IEEEtran}
\bibliography{sigproc}

\end{document}